\documentclass[draft]{agujournal2019}
\usepackage{url} 
\usepackage{lineno}
\usepackage[inline]{trackchanges} 
\usepackage{epstopdf}
\usepackage{soul}
\draftfalse

\journalname{Space Weather}

\begin{document}

%
%


\title{Analysis of Large Geomagnetically Induced Currents During the 7-8 September 2017 Storm: Geoelectric Field Mapping}




\authors{Anna Wawrzaszek\affil{1}, Agnieszka Gil\affil{1,2}, Renata Modzelewska\affil{2}, Bruce T. Tsurutani\affil{3} and Roman Wawrzaszek\affil{1}}
\affiliation{1}{Space Research Centre, Polish Academy of Sciences, Bartycka Str. 18A, 00-716 Warsaw, Poland}
\affiliation{2}{Faculty of Exact and Natural Sciences, Institute of Mathematics, Siedlce University, Konarskiego Str. 2, 08-110 Siedlce, Poland}

\affiliation{3}{Jet Propulsion Laboratory, California Institute of Technology, Pasadena, California, 91109, USA}



\correspondingauthor{Anna Wawrzaszek}{anna.wawrzaszek@cbk.waw.pl}





\begin{keypoints}
\item A new 10s resolution GeoElectric Dynamic Mapping (GEDMap) procedure was developed to study Geomagnetically Induced Currents (GICs).
\item The modeling results performed using the geomagnetic field mapping and M\"ants\"al\"a GIC measurements are in excellent agreement.
\item Rapid and strong evolution of the geoelectric fields occurred at and near M\"ants\"al\"a during the two largest GIC events of the magnetic storm.

\end{keypoints}

    
%
%

%
%


\begin{abstract}

High temporal and high spatial resolution geoelectric field models of two M\"ants\"al\"a, Finnish pipeline GIC intervals that occurred within the 7-8 September, 2017 geomagnetic storm have been made. The geomagnetic measurements with 10 s sampling rate of 28 IMAGE ground magnetometers distributed over the north Europe (from $52.07^\circ$ to $69.76^\circ$ latitude) are the bases for the study. A GeoElectric Dynamic Mapping (GEDMap) code was developed for this task. GEDMap considers 4 different methods of interpolation and allows a grid of $0.05^\circ$ (lat.)$\times 0.2^\circ$ (lon.) spatial scale resolution. The geoelectric field dynamic mapping output gives both spatial and temporal variations of the magnitude and direction of fields. The GEDMap results show very rapid and strong variability of geoelectric field and the extremely localized peak enhancements. The magnitude of geoelectric fields over M\"ants\"al\"a at the time of the two GIC peaks were 279.7 mV/km and 336.9 mV/km. The comparison of the GIC measurements in M\"ants\"al\"a and our modeling results show very good agreement with a correlation coefficient higher than 0.8. It is found that the auroral electrojet geoelectric field has very rapid changes in both magnitude and orientation causing the GICs. It is also shown that the electrojet is not simply oriented in the east-west direction. It is possible that even higher time resolution base magnetometer data of 1 s will yield even more structure, so this will be our next effort.

\end{abstract}

\section*{Plain Language Summary}
The Sun is an active star, the state of which has a strong influence on conditions on the Earth, in particular leading to increased fluctuations in the geomagnetic field and formation of strong ground electric fields (GEFs). One of the most important consequences of exceptional high levels of GEF is the occurrence of  geomagnetically induced currents (GICs), which are particularly dangerous for electrical infrastructure and the increase the number of grid failures. In the frame of this work, to better understand causes of GICs  during a strong GIC event on 7-8 September, 2017,  we have developed a new procedure to perform systematic computation of geoelectric field at latitudes from $52.07^\circ$ to $69.76^\circ$. Next, to reveal  spatio-temporal evolution of GEF (its magnitude and direction) and to give a global perspective we propose a robust method of construction of maps. Results show rapid and strong evolution of geoelectric field.

\section{Introduction}
\label{sec:1}
Geomagnetically induced currents (GICs) are intense, low-frequency currents induced in large conductive systems like power lines and pipelines, during space weather events \cite<e.g.>[]{Laket21}. GICs, depending on the solar wind and the Earth’s electrical conductivity structure, were analyzed extensively for many years \cite<e.g.>[and references therein] {Kap96,Botet98,Pir02,Pulet15,Kel19,Laket21}, based on various physical characteristics: solar wind parameters, electrojet indices, magnetospheric drivers, as well as systematically developed models. \\
Consideration of the occurrence of GICs in power grids was mostly focused on the impact of intense sudden perturbations of the geomagnetic field, $B$. Variations of $dB/dt$ were often used in GIC evaluations for comparisons between different storms, for producing a proxy GIC index \cite<e.g.>[]{Maret11}.  A large number of studies focused on the realistic reproduction of GICs from the geomagnetic data measured by IMAGE (International Monitor for Auroral Geomagnetic Effects) \cite{Dimmock2019, Dimet20, Dimmock2021}, SuperMAG \cite{Hajra2020,2020Despirak,Tsurutani2021, Cliet21}, or INTERMAGNET (International Real-time Magnetic Observatory Network)  \cite<e.g.>[]{2020Piersanti} stations. 
IMAGE \cite<e.g.>[]{Viljanen95}, INTERMAGNET \cite<e.g.>[]{Jankowski}, and SuperMAG \cite<e.g.>[]{Gjerloev} are the international networks of magnetic observatories.
\\
The importance of the regional variability of geomagnetic disturbances was very often emphasized \cite<e.g.>[]{VilPir17,Dimet20,2021Boteler,2021Svanda, 2021Torta}. Nevertheless, the sparse distribution of magnetometer stations demanded the application of the interpolation of the geomagnetic field. A review of the literature shows that various methods have already been applied in this context \cite<e.g.>[]{McLet10,2017Torta}. We can mention both purely mathematical methods, such as a nearest neighbor, linear interpolation using Delaunay triangulation \cite{Del34} and those with the physical background, such as magnetic scalar potential \cite{Duzet97} and one of the most  widely used Spherical Elementary Current Systems (SECS) interpolation scheme \cite<e.g.>[]{Amm97,AmmVil99}. Studies indicated that SECS has been shown to give good results at high latitudes, however, this technique was not found to be particularly accurate for $B$ interpolation purposes at middle latitudes \cite<e.g.>[]{McLet10,2017Torta}. In particular, \citeA{2017Torta} performed a systematic comparison of four mentioned techniques for region of Spain and showed that using a nearest neighbor technique or a magnetic scalar potential method achieved better results than by using the SECS interpolation scheme. These results seem to be in agreement with analysis in the UK with a similar sparse magnetometers distribution \cite{McLet10}. A more detailed discussion about application of SECS interpolation scheme beyond high latitudes can be found in e.g. \citeA{2017Torta}. 
\\
The review of the literature devoted to geomagnetically induced currents reveals that various drivers of GICs were indicated \cite{Tsurutani2021, Oliet18}. In particular, drivers of intense GICs were often associated with large storm impulsive events such as coronal mass ejections (CMEs) and their upstream shocks and sheaths \cite<e.g.>[]{87GT, 93Gosling, 1997T}. Fluctuations different from regular oscillations of the geomagnetic field, or geomagnetic pulsations, have also been identified as possible drivers of GICs \cite{Heyet2020}.  At mid-latitude and low-latitude, large GICs have been related to storm sudden commencements and sudden impulses, rather than substorms \cite<e.g.>[]{Maret12}.\\
Nevertheless, comprehensive studies confirm the complexity of geophysical phenomena responsible for GICs, both from the ionospheric and magnetospheric current systems point of view. Thus, GIC causes during some selected events are still not fully explained. For example, 7--8 September 2017 geomagnetic storm was one of the largest storms of the 24th solar cycle \cite{Dimmock2019, Beget21} with two largest geomagnetically induced currents episodes (GIC$\sim 30$ A). Despite numerous and extensive analyses of space weather activity on 7--8 September 2017, some recent papers pointed out several, still unexplained, key features of GIC observed during this geomagnetic storm \cite<e.g.>[]{Hajra2020, Cliet21}. In particular, \citeA{Hajra2020} stated that the largest GIC of this interval cannot be associated with any large substorm or any solar wind feature. \citeA{Cliet21} asked about the up-stream drivers of the GIC event, the scale-sizes of the driving mechanisms, why and which magnetic local time sector was important for large GIC occurrence.\\
It is worth underlining also that most studies devoted to the September 2017 storm, considered the SECS method and determined ionospheric equivalent currents focusing on latitudes higher than $60^\circ$ \cite<e.g.>[]{Dimmock2019}. Regions located at latitudies lower than $60^\circ$ were considered less extensively \cite<e.g.>[]{22Krug} and further analysis are still needed, especially in the light of recent studies devoted to transmission line failures in Poland \cite{Gilet21}. Moreover, despite many sophisticated methods deployed to evaluate the geoelectric field over large regions \cite<e.g.>[]{Velet16} it is worth using simpler approaches, for example \cite{Pieet19}, where $E$ was first computed and then spatial maps were evaluated through a spherical harmonics interpolation.\\
Therefore, in the frame of this work, we continue systematic studies of large GIC cases occurring during 7--8 September 2017 geomagnetic storm. For this purpose, in the first step, based on a great collection of geomagnetic field data sets from $28$ IMAGE magnetometers located at latitudes from $52^\circ$ to $70^\circ$, we perform a systematic computation of geoelectric fields and their variations. Then, having values of geoelectric field determined for the positions of magnetometers and applying  the natural neighbor interpolation, we construct spatial maps of the geoelectric field. This type of interpolation has not been used previously; this method has been compared with other interpolation methods and was shown to be more effective for the considered data. Prepared  GeoElectric Dynamic Mapping (GEDMap) states efficient and informative global image of spatial-temporal evolution of geoelectric field  around the periods when the two largest GICs were registered in M\"ants\"al\"a, Finland. \\
Moreover, proposed geoelectric mapping allows for determination of GICs in selected locations, not covered by magnetometers. Results reveal a strong variability of geoelectric field and the extreme localized peak enhancements during the intense 7--8 September 2017 geomagnetic storm. The directional evolution of the strongest values of the field, a rapid change of the orientation of geoelectric field, as well as to differentiate the two largest GICs events seems to be important for better understanding causes of GICs.
The data used in analysis are described in Section~\ref{sec:2}. In Section \ref{sec:3} we briefly discuss the methodology applied to compute geoelectric field and GICs. Section~\ref{sec:4} presents the results of the analysis in the form of maps.  A summary is provided in Section \ref{sec:5}. Additional detailed animations are presented in Supporting Information.

\section{Data}
\label{sec:2}
\subsection{M\"ants\"al\"a Finland Pipeline Measurements}
\begin{figure}[h]
\centering
\includegraphics[width=0.96\columnwidth]{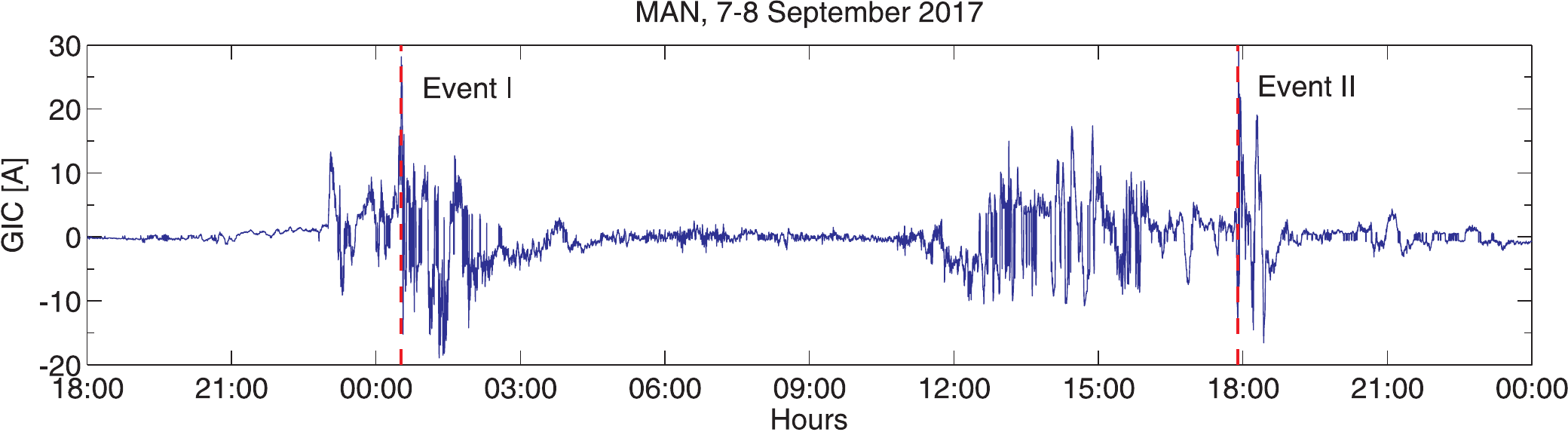}
\caption{\small GIC along the Finnish natural gas pipeline at the M\"ants\"al\"a station on 7--8 September 2017 geomagnetic storm (between 18:00 UT on September 7 and 23:59 UT on September 8, 2017). The Event I with GIC=28.18 A at 00:31:20 UT and the Event II with GIC=30.41 A at 17:54:40 UT have been marked by red dashed lines.}
\label{fig:1}
\end{figure}

GIC recordings in the Finnish natural gas pipeline near M\"ants\"al\"a located at subauroral latitudes (57.9$^\circ$N geomagnetic latitude; 60.6$^\circ$N geographic latitude, 25.2$^\circ$E geographic longitude) \cite{Vilet06}.
Figure \ref{fig:1} shows measurements during 7--8 September 2017 geomagnetic storm, when the two largest GIC events (GIC$>20$ A) were observed. Namely, Event I with GIC=28.18 A at 00:31:20 UT, Event II with GIC=30.41 A at 17:54:40 UT (denoted by red dashed lines), as well as several smaller events when GIC $\sim 10$ A.

\subsection{Geomagnetic Field Data}
\label{subsec:2.1}

We consider here data of geomagnetic field components registered by 28 magnetometers being the part of IMAGE network (\url{http://space.fmi.fi/image/}). 
The criterion for the data was the resolution of 10 s collected by IMAGE and the existence of a conductivity/resistivity model to calculate the geoelectric field.
Details of the stations taken into account, as geographic longitudes and latitudes, as well as their Corrected Geomagnetic Coordinates (CGM), are presented in Table \ref{table:1}.

\begin{table}[h]
\caption{Summary of observatory data used in the analysis}
\centering
\label{table:1}
\begin{tabular}{cccccccc}
\hline
Code & Name & Geogr. Lat  & Geogr.   Lon & CGM Lat     & CGM  Lon    & No of model \\
                      &                       & [$^\circ$] & [$^\circ$] &  [$^\circ$] &  [$^\circ$] &  \\\hline
KEV                   & Kevo                  & 69.76   & 27.01   & 66.32   & 109.24  & 26                                                   \\
TRO                   & Tromsø                & 69.66   & 18.94   & 66.64   & 102.9   & 26                                                    \\
KIL                   & Kilpisj\"arvi           & 69.06   & 20.77   & 65.94   & 103.8   & 26                                                    \\
IVA                   & Ivalo                 & 68.56   & 27.29   & 65.1    & 108.57  & 26                                                   \\
ABK                   & Abisko                & 68.35   & 18.82   & 65.3    & 101.75  & 26                                                    \\
MUO                   & Muonio                & 68.02   & 23.53   & 64.72   & 105.22  & 26                                                    \\
KIR                   & Kiruna                & 67.84   & 20.42   & 64.69   & 102.64  & 26                                                   \\
RST                   & Røst                  & 67.52   & 12.09   & 64.88   & 95.9    & 25                                                   \\
SOD                   & Sodankyl\"a             & 67.37   & 26.63   & 63.92   & 107.26  & 24                                                   \\
PEL                   & Pello                 & 66.9    & 24.08   & 63.55   & 104.92  & 26                                                   \\
JCK                   & J\"ackvik               & 66.4    & 16.98   & 63.51   & 98.31   & 26                                                  \\
DON                   & Dønna                 & 66.11   & 12.5    & 63.38   & 95.23   & 25                                                  \\
RAN                   & Ranua                 & 65.9    & 26.41   & 62.09   & 105.91  & 24                                                 \\
RVK                   & Rørvik                & 64.94   & 10.98   & 62.23   & 93.31   & 25                                                \\
LYC                   & Lycksele              & 64.61   & 18.75   & 61.44   & 99.29   & 25                                                    \\
OUJ                   & Ouluj\"arvi             & 64.52   & 27.23   & 60.99   & 106.14  & 24                                                    \\
MEK                   & Mekrij\"arvi            & 62.77   & 30.97   & 59.1    & 108.45  & 24                                                    \\
HAN                   & Hankasalmi            & 62.25   & 26.6    & 58.69   & 104.54  & 25                                                    \\
DOB                   & Dombås                & 62.07   & 9.11    & 59.29   & 90.2    & 24                                                  \\
SOL                   & Solund                & 61.08   & 4.84    & 58.53   & 86.26   & 24                                                  \\
NUR                   & Nurmij\"arvi            & 60.5    & 24.65   & 56.89   & 102.18  & 25                                                  \\
UPS                   & Uppsala               & 59.9    & 17.35   & 56.51   & 95.84   & 24                                                   \\
KAR                   & Karmøy                & 59.21   & 5.24    & 56.43   & 85.67   & 24                                                   \\
TAR                   & Tartu                 & 58.26   & 26.46   & 54.47   & 102.89  & 41                                                    \\
BRZ                   & Birzai                & 56.17   & 24.86   & 52.3    & 100.81  & 1                                                    \\
SUW                   & Suwałki               & 54.01   & 23.18   & 49.97   & 98.7    & 40                                                  \\
WNG                   & Wingst                & 53.74   & 9.07    & 50.01   & 86.65   & 12                                                  \\
NGK                   & Niemegk               & 52.07   & 12.68   & 47.96   & 89.13   & 14                           \\
\hline
\end{tabular}
\end{table}

\subsection{Geomagnetic Indices}
\begin{figure}[!h]
\centering
\includegraphics[width=0.75\columnwidth]{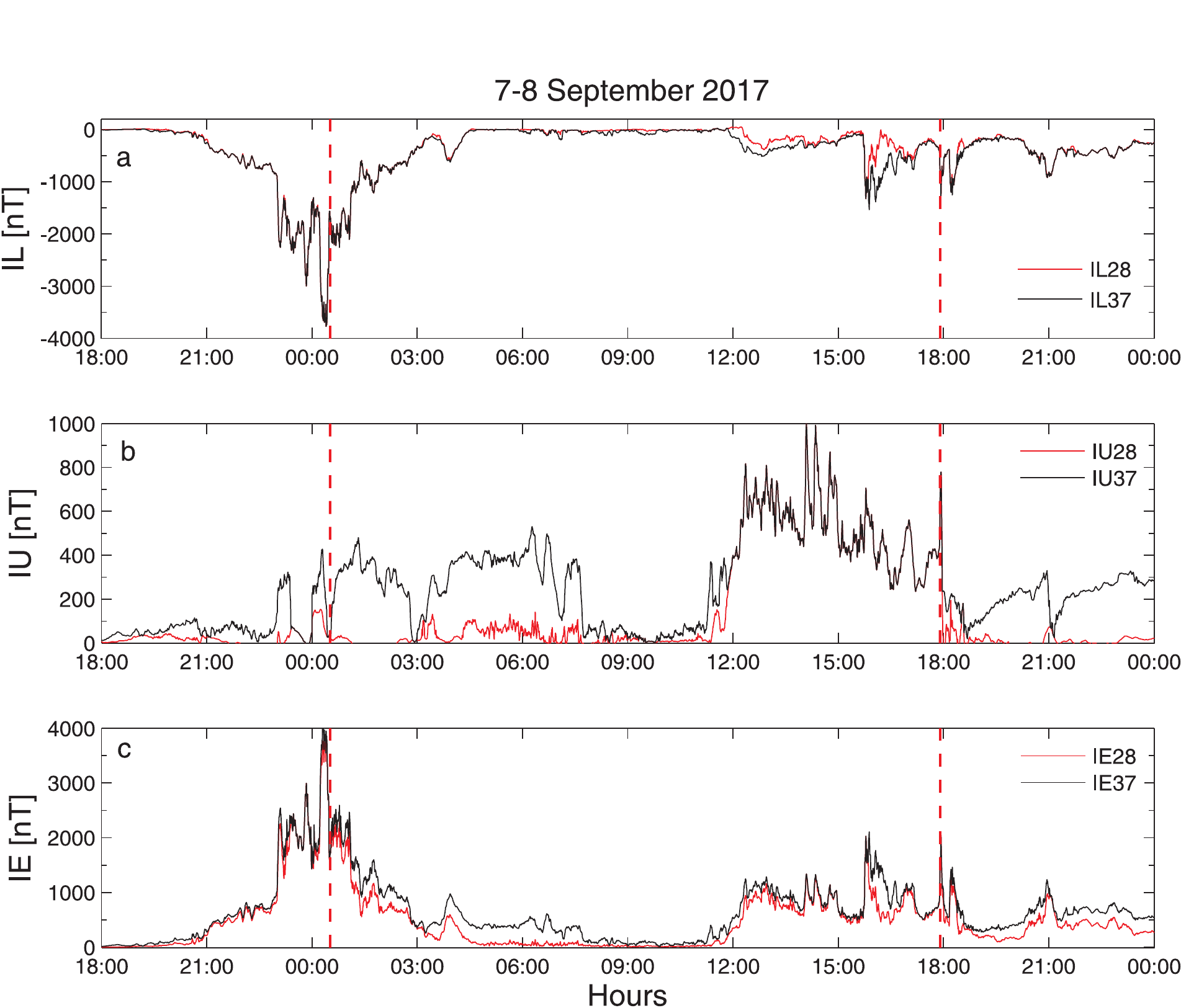}
\vspace{5pt}
\caption{\small IMAGE derived geomagnetic indices IL (a), IU (b), and IE (c) between 18:00 UT on September 7 and 23:59 UT on September 8, 2017 (time step 10 s). Black lines represent the indices generated for all available IMAGE stations (in summary 37), while the red lines correspond to the case when only 28 stations listed in Table \ref{table:1} were considered. Red dashed lines indicate the moment of Event I at 00:31:20 UT and Event II at 17:54:40 UT, respectively.}
\label{fig:IE}
\end{figure}

We also study electrojet indicators from IMAGE. These are basic measures of the total westward and eastward currents, which cross the magnetometer grid. They are defined similarly to the standard AU, AL, and AE indicesc \cite{Davis66}. The AE index is derived from geomagnetic variations in the horizontal component observed at selected (10--13) observatories along the auroral zone in the northern hemisphere. The AU and AL indices are respectively defined by the largest (upper) and the smallest (lower) values. Correspondingly, IU-index is estimated as the maximal variation value ${\Delta B_X(t)}$ of the geographical north magnetic field components ($IU(t)=\max({ \Delta B_X(t)})$), measured at all selected stations. The variation is obtained in relation to the quiet period, based on the measurements of the selected observatories. In the analogous way IL-index is estimated as the minimal variation, ($IL(t)=\min({ \Delta B_X(t)})$), and finally IE-index is a difference between IU and IL indices: $IE(t) = IU(t)-IL(t)$. For 7--8 September storm indices are displayed in the Figure \ref{fig:IE}. Here, the quiet time is defined as a three-hour interval before the considered storm.
We present these indices based on all available IMAGE stations, i.e., 37 (represented by black lines) in comparison with the case of 28 stations (red lines), listed in Table \ref{table:1}, for which models with values of thicknesses and resistivities are available \cite{Adaet12}.

It is worth underlining that Event I denoted in Fig. \ref{fig:IE}  by red dashed line at 00:31:20 UT was caused by the fast interplanetary shock observed at $\sim$23:00 UT on 7 September, related to a fast halo CME on 6 September at 12:24 UT of class X9.3 flare (SOHO LASCO CDAW catalog, \url{cdaw.gsfc.nasa.gov}). 
The magnetic cloud detected between $\sim$20:24 to $\sim$23:02 UT on September 7 was characterized by low solar wind temperature ($\sim$0.16$\times$ 105 K) and low plasma $\beta$ ($\sim$0.06) with southward heliospheric magnetic field (HMF) $B_z$ peak of $\sim$-10 nT at $\sim$21:06 UT on September 7. Next the fast shock was followed by highly compressed sheath region that extended from $\sim$23:02 UT on September 7 to $\sim$11:31 UT on September 8 with a peak southward $B_z$ of $\sim$-31 nT at $\sim$23:31 UT on September 7 \cite<e.g.>[]{Hajra2020}. 
The passage of the shock over the CME structure, strong southward $B_z$, and the strongly compressed sheath is the most probable reason of the Event I strong geomagnetic storm. Geomagnetic storm 7--8  September was shown \cite{Hajra2020} as ‘three-step feature’ in Sym-H index with resolution of 1 min and cannot be observed in hourly Dst-index.\\
Moreover, Event I 
was followed by peak IE and IL IMAGE indices of 3813.3 nT and -3772.7 nT, respectively at $\sim$00:24 UT on September 8 (Figure \ref{fig:IE}). Almost at the same time the Nurmij\"arvi magnetometer recorded a significant magnetic depression and large $dB_X/dt$ fluctuations at this time \cite{Dimmock2019}. This event took place over the passage of slowly developing supersubstorm of long-duration $\sim$3 hr 49 minutes from 7 September  $\sim$23:02 UT to 8 September $\sim$02:51 UT with peak SME and SML intensities (SuperMAG, \url{https://supermag.jhuapl.edu/}) of 4464 and -3712 nT, respectively, at $\sim$00:24 UT on September 8 \cite{Hajra2020}. It is worth mentioning that there is an interval of $\sim$1 hr from the minimum of $B_z$ ($\sim$-31 nT at $\sim$23:31 UT on 7 September) and the maximum of GIC (28.18A at 0:31:20 UT on 8 September) observed in M\"ants\"al\"a.\\
The Event II appeared around 17:54:40 UT, on September 8, during the recovery phase of extremely intense substorm \cite{Hajra2020}. During that time a magnetic cloud was noticed from around 11:31 UT to 18:00 UT \cite<e.g.>[]{Hajra2020} with peak southwardly directed HMF $B_z$ component value being -17 nT and solar wind speed 790 km/s. This southward $B_z$ component caused an intense magnetic storm, starting around 11:26 UT with peak Sym-H index value -115 nT at $\sim$ 13:56 UT \cite<e.g.>[]{Hajra2021}. At the main phase of this intense magnetic storm revealed another supersubstorm with SuperMAG indices \cite<e.g.>[]{Newell11} SML and SME values -2642 nT and 4330 nT, respectively \cite{Hajra2021}. 
IMAGE geomagnetic indices IL and IE during this supersubstorm had the extremal values at 15:52:50 UT, being equal -1534.5 nT and 2112.3 nT, respectively and IU index reached local maximum earlier, at 14:21:00 at the level of 992.6 nT (Figure \ref{fig:IE}). It is worth mentioning that Event II occurred at end of the strongest growth in the IU index (Figure \ref{fig:IE}). 
Although, it is worth bearing in mind that 'There are no obvious 1-to-1 relationships between GIC events and substorms' \cite{Tsurutani2021}.
Moreover, Kp-index value around the Event II was 8 in 12:00-15:00 UT and 7+ in 15:00-18:00. \\

This 7--8 September geomagnetic storm and causative CME was possibly related to a moderate solar flare of M2.9 intensity, although around the time of Event II there was only the solar flare of M2.9 class, lasting from 15:09 UT to 16:04 UT, September 8, with peak time at 15:47 with source located in S09W63 of active region no. 12673.\\
Since the global average total electron content (TEC) characterizing the ionospheric conditions exhibits firm sensitivity to enhanced solar activity \cite<e.g.>{Nikitina2022}, it is worth mentioning that TEC at the middle and high latitudes showed during 7-8 September storm a strong  hemispheric asymmetry, as well as a long -duration recovery of topside TEC with respect to the pre-storm state \cite{Tsuet05,Jimoh2019}.\\

\section{Methodology}
\label{sec:3}

\subsection{Calculation of the geoelectric field}
\label{sub:m1}

In order to estimate geoelectric field $E$ from measured 10 s geomagnetic field data $B$ we applied a 1-D layered conductivity Earth model \cite{Botet19,BotPir19}. Earth conductivity varies in all directions, however, the biggest variation of conductivity is with the depth \cite{BotPir19}. Therefore, Earth is often represented by a 1-D model \cite<e.g.>[]{BotPir19,Khuet20}, in the frame of which we have $N$ layers, each specified by conductivity $\sigma_{n}$ and thickness $l_{n}$ ($n=1,...,N$). Then, layered case of the transfer function $K$ (in the frequency domain $f$) is expressed by the following recursive formula \cite{Wea94,BotPir19}:
\begin{equation}
K_{n} =\eta_{n}\frac{K_{n+1}(1+\mathrm{e}^{-2 k_{n}l_{n}})+\eta_{n}(1-\mathrm{e}^{-2 k_{n}l_{n}})}{K_{n+1}(1-\mathrm{e}^{-2 k_{n}l_{n}})+\eta_{n}(1+\mathrm{e}^{-2 k_{n}l_{n}})}
\label{e:m:1}
\end{equation}
where $K_{n}$ is the ratio of $E$ to $B$ at the top surface of layer $n$, while $K_{n+1}$ at the top surface of the underlying layer $n+1$, $\eta_{n}=\frac{i 2\pi f}{k_{n}}$, $k_{n}=\sqrt{i 2\pi f \mu_{0}\sigma_{n}}$ and $\mu_{0}=4\pi10^{-7}$ Hm$^{-1}$ \cite{Botet19}.
The initial value in Equation (\ref{e:m:1}) corresponds to the case when the layer $n=N$ is a uniform half-space and $K_{N}=\sqrt{\frac{i 2\pi f}{\mu_{0}\sigma_{N}}}$. The final value $K_{1}$ ($n=1$) is the transfer function relating $E$ and $B$ at the Earth's surface \cite{TriBot02, Botet19}.
In the frame of the paper, to perform analysis of data from 28 stations located at different latitudes and longitudes, we applied various 1-D Earth resistivity models, as listed in Table \ref{table:1}. These models, described in details in \cite{Adaet12} vary in the number of layers, values of thicknesses ($l$), and resistivities ($1/\sigma$), and correspond to local conditions around the observatory stations. 
In the next step of the analysis, the geomagnetic field $\{B_X,B_Y\}$ was decomposed into its frequency components $\{B_X(f),B_Y(f)\}$ and multiplied by the corresponding transfer-function values,  namely $E_{X}(f)=K(f)B_{Y}(f)$ and $E_Y(f)=-K(f)B_{X}(f)$, where $E_{X}(f)$ and $E_{Y}(f)$ denote geoelectric field frequency components. In the final step, we employed the inverse Fourier transform to obtain the value of a geoelectric field in the time domain $[E(t)]$ for both northward $[E_X]$ and eastward $[E_Y]$ components \cite{Bot12,BotPir19}. 
Please note, that the Earth's internal structure can have a complex 3-D distribution of electrical resistivity. 
However, various studies \cite{Vilet13,Vilet14,Gilet21} confirmed that a 1-D model can still be treated a useful first-order approximation for modeling gross resistivity structure and its effect on surface electric fields. Moreover, it is worth noting, that the 1-D model is fast and accurate at a single location \cite{Beget21}, which is the case for this study.
Since we use a local conditions of 1-D model, with a relatively dense net of magnetometers, one may assume that geoelectric field can be estimated only from geomagnetic field variations, without ionospheric current considerations \cite<compare, e.g.>[]{2021Svanda}.

\subsection{Spatial Interpolation of Geoelectric Field}
\label{sub:m2}

Various methods of interpolation can be applied to reconstruct a surface from irregularly distributed sample points. Interpolation may or may not require any physical or statistical assumptions about properties and behaviour of interpolated parameter over the surface where its value is desired. In the present analysis, we focused on 2-D interpolation methods which do not require those assumptions. More precisely, we considered how useful in the context of geoelectric field maps preparation can be four mathematical methods: the nearest neighbor, the linear interpolation, the cubic spline interpolation \cite{DeBoor78} and the natural neighbour interpolation \cite{Sib81}. \\

\begin{figure}[!h]
  \centering
\includegraphics[width=0.9\columnwidth]{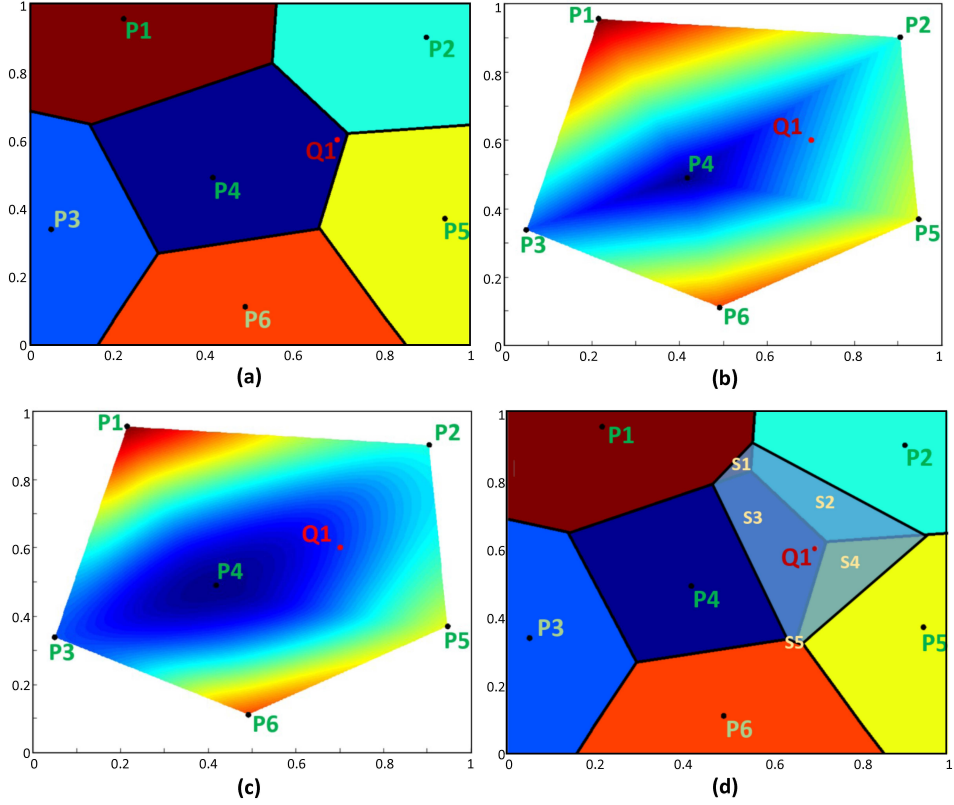}
  \caption{Illustration of 2D interpolation methods. Each diagram shows interpolation of data picked in point $Q1$ if there are given or measured values in locations from $P1$ to $P6$. On the diagrams the same color represents the same signal level. Value in location $Q1$ is interpolated depending on the following methods: a) the nearest neighbor method - value in point $Q1$ is equal to value in $P4$;  b) the linear interpolation – value in $Q1$ is triangulated from 3 known values on a triangle defined by $P2$, $P4$, $P5$ vertices, which include $Q1$; c) the cubic spline interpolation – value in $Q1$ is based on a cubic interpolation of the values at neighboring points in each respective dimension; d) the natural neighbor - value in $Q1$ is a function of values in $P1$, $P2$, $P4$, $P5$ and $P6$ weighted by overlapping areas $S1$,...,$S5$.}
  \label{fig:in_met_comp}
  \end{figure}

The simplest of considered interpolation methods is the nearest neighbor technique \cite<e.g.>{Han13}. In this method one assumes that the value of parameter assigned to particular location is equal to the real measurement performed in the nearest station, taking into account direct distance. This method is not adequate for analysed phenomena due to its discontinuity, lack of smoothness and differentiability. Nevertheless, it is a simple method and may state a reference point for more advanced ones.\\
The second of considered methods is the linear interpolation, in the frame of which the scattered data set is first triangulated using a Delaunay triangulation \cite{Del34}. The interpolated value at a query point $Q1$ is then derived from the values of the vertices of the triangle that enclose the point (see Figure \ref{fig:in_met_comp}b for illustration). As a result, the whole interpolated surface is $C^0$ continuous.\\
The cubic spline interpolation algorithm is convolution-based interpolation described in details in \cite{DeBoor78,Key81}. In a frame of this algorithm one assumes that the given points are joined by a cubic polynomial. To find the interpolating function, the determination of the four cubic polynomial coefficients for each of the cubic function is required, where in the case of $n$ points, there are $n-1$ cubic functions. To determine unknown parameters of polynomial function it is assumed that function must be met in known points, as well as both the first and the second derivatives of all polynomials are identical in the points where they touch their adjacent polynomial. For 2-D problems, the interpolation is done for one direction followed by another. As a result, the interpolated surface is smooth, continuous and differentiable. The effect of the cubic spline interpretation algorithm has been schematically shown in Figure \ref{fig:in_met_comp}c.\\
The natural neighbor interpolation proposed by \citeA{Sib81} is sourced in the nearest neighbor and partially solves disadvantages of this classical method. Natural neighbor method produces a surface which is $C^1$ continuous (except at the sample points), provides good continuity for slope, smoothness and visually appealing results. In this case, picked point $Q1$ is considered as an artificial measurement point to determine area which would be associated with this point in the classical nearest neighbor method. Interpolated value is calculated then as an average of overlapped the nearest neighbor regions weighted by size of areas overlapping with regions originally derived from the nearest neighbor method (see Figure \ref{fig:in_met_comp}d ). It is worth to underline that natural  neighbor interpolation is parameter free, creates a smooth surface free of any discontinuities, requires no statistical assumptions and can be applied to small datasets \cite<e.g.>[]{Samet95}.
\clearpage

\subsection{GIC calculation}
\label{sub:m3}

In a uniform electric field, GICs are computed using the formula $GIC=a\cdot E_{X}+b\cdot E_{Y}$, where $E_{X},E_{Y}$ are the local geoelectric field components, while constants $a$ and $b$ depend on the power grid parameters such as the resistances or the network topology \cite{PL85}. Here, we used various pairs ($a$, $b$)[Akm/V] for particular locations, available in the literature, \cite<e.g.>[]{Pulkkinen01,W2008,Zhang12,Dimmock2021}. The details of the ($a$, $b$) used are gathered in the Table \ref{table:ab}. For each of the mentioned locations we have estimated the 10s data of the  geoelectric field and then calculated the GIC values for each of them.
\begin{table}[h]
\caption{Summary of the pairs ($a$, $b$) used in the GIC calculations}
\centering
\label{table:ab}
\begin{tabular}{cccccc}
\hline
a  & b  & References & Place &  Geogr. Lat  & Geogr. Lon     \\
$[$Akm/V$]$       & $[$Akm/V$]$ & &       &  [$^\circ$] & [$^\circ$] \\
\hline
9.6        & 5.9     & \citeA{Zhang12}& Pirttikoski  & 66.3     & 27.1  \\ 
-36   & 180     & \citeA{Pulkkinen01} & Imatra       & 61.2     & 28.8     \\
24         & 190       & \citeA{Pulkkinen01} & Kouvola        & 60.9     & 26.3\\
-70.0     & 88.0        & \citeA{Pulkkinen01}& M\"ants\"al\"a & 60.7     & 25.0     \\
    &         & &(MAN) &    &   \\
-74    & 43         & \citeA{Pulkkinen01} & H\"ameenlinna                & 61.0     & 24.2  \\
-62.3     & 133.2      & \citeA{W2008}& Simpevarp-2    & 57.4     & 16.6\\
\hline
\end{tabular}
\end{table}
\section{Results and discussion}
\label{sec:4}
To perform a systematic analysis of spatial-temporal variability of geoelectric field during the selected 7-8 September, 2017 geomagnetic storm and for considered region of the north Europe, a  GeoElectric Dynamic Mapping (GEDMap) code has been prepared. GEDMap keeps the time resolution of input geomagnetic field time series (Sect. \ref{subsec:2.1}), applies 1-D layered conductivity Earth models (Sect. \ref{sub:m1}) and considers 4 different methods of interpolation (Sect. \ref{sub:m2}), allowing in final step for $E$ maps preparation with a grid of $0.05^\circ$ (lat.)$\times 0.2^\circ$ (lon.) spatial scale resolution. In the further parts of this Section various outputs of GEDMap has been shown and discussed.

\subsection{Selection of method of interpolation}
\label{sec:4:1}

\begin{figure}[!h]
  \centering
\includegraphics[width=0.80\columnwidth]{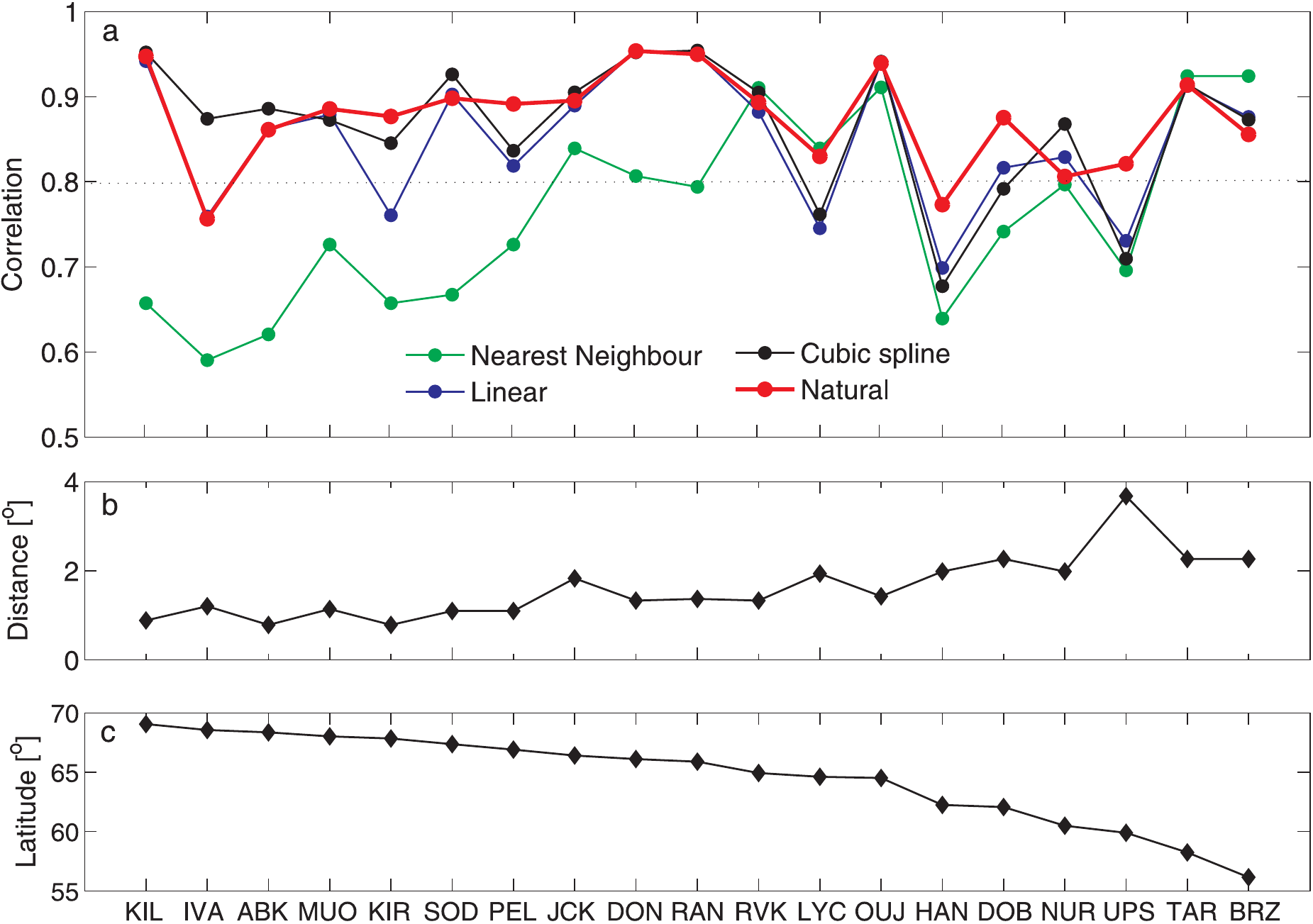}
  \caption{ (a) Correlation values between $E$ determined in measurement point and $E$ obtained as a result of interpolation performed by using nearest neighbor (green line), linear (blue line), cubic spline (black line) and natural neighbor (red line) techniques; (b) distance of the selected station to nearest one in $^\circ$ - great circle arc connecting 2 points on a surface. Here $1^\circ\sim 111$ km ; (c) geographic latitudes of the 19  considered as verification points stations (Table  \ref{table:1}). }
  \label{fig:corr_stat}
  \end{figure}

Accuracy and usefulness of considered four methods of interpolation (Sect. \ref{sub:m2}), has been identified using procedure in which real data registered by station in particular location was compared with interpolated value calculated for the same location but excluding the station from the set of known data points during interpolation. This procedure was performed for all stations in considered region.\\
Figure \ref{fig:corr_stat} shows the summary of the performed comparison. In particular, Figure \ref{fig:corr_stat}a presents correlation values between $E$ determined in measurement point (at 19 stations from 28 listed in Table \ref{table:1}, indicated on x axis) and interpolated by using nearest neighbor (green line), linear (blue line), cubic spline (black line) and natural neighbor interpolation (red line). It is worth noting that the correlation was computed using 10 s resolution data over the full considered time frame, namely, between 18:00 UT on September 7 and 23:59 UT on September 8, 2017. Figure \ref{fig:corr_stat}b shows the distance of the selected station to nearest one, expressed in $^\circ$ - great circle arc connecting 2 points on a surface, where one degree corresponds to $\sim 111$ km. Figure \ref{fig:corr_stat}a  indicates the  geographic latitude on which the considered  measurement station and verification point are located. It is worth stressing that for about $13$ first stations located on higher latitudes the distance to the nearest stations is smaller than for those placed on latitudes below $65^\circ$ where, in general, more irregular and sparse spatial coverage of observatory-grade magnetometers is observed.

Systematic analysis of Figure \ref{fig:corr_stat} reveals different level of usefulness of interpolation methods, which directly depends on the considered stations. The lowest correlation between a given and interpolated  $E$ is observed for cases when the nearest neighbor interpolation has been applied. Unexpectedly, the worst results are observed for stations densely distributed at higher latitudes.
For the rest of three methods (linear, cubic spline and natural neighbour) the worst results are observed for stations sparsely located on latitudes below $65^\circ$. The lowest correlation is observed for stations LYC, HAN and UPS. Please note that for station UPS the distance to the nearest station (Figure \ref{fig:corr_stat}b) is the biggest, around $4^\circ$, (440 km).
The most adequate method of interpolation suggested by Figure \ref{fig:corr_stat}c seems to be the natural neighbor (red line), results of which present the highest level of correlation between $E$ determined and interpolated for the most of considered locations (stations). In the latter part of the paper, we will focus on application of natural neighbor method. 
Performed in this section analysis states the basic check of credibility for adequate approximation of physical phenomena that we are analyzing. One should note, however, that the interpolation itself does not reflect phenomena of propagation of geoelectric field in a physical sense. Nevertheless, it can be treated as a first approximation of the considered  process behaviour to analyze its spatial evolution and rough properties.

\subsection{Geoelectric Field Mapping}
\begin{figure}
  \centering
 \includegraphics[width=0.8\columnwidth]{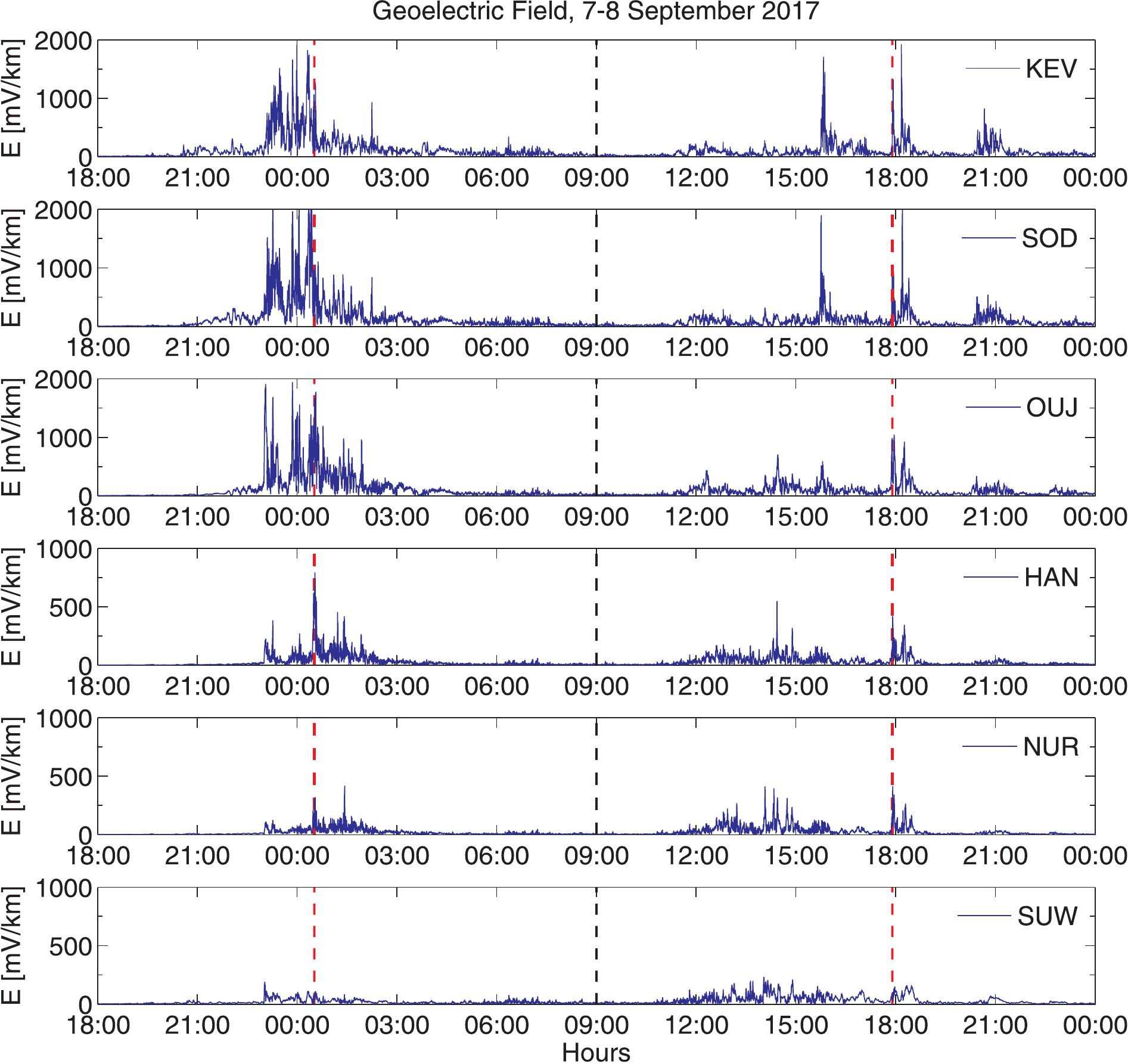}
  \caption{\small Time variation of geoelectric field $E=\sqrt{E_X^2+E_Y^2}$ between 18:00 UT on September 7 and 23:59 UT on September 8, 2017 (time step 10 s) for six selected stations: KEV, SOD, OUJ, HAN, NUR, SUW.}
  \label{fig:E}
  \end{figure}

Figure \ref{fig:E} presents the magnitude of the modelled geoelectric field $E=\sqrt{E_X^2+E_Y^2}$ between 18:00 UT on September 7 and 23:59 UT on September 8, 2017, for six selected stations: KEV, SOD, OUJ, HAN, NUR, SUW (listed in Table \ref{table:1}). Please note that the y-axis from the bottom three panels shows half the range of the top three. One can see the expected increase of $|E|$ for Event I and Event II, denoted by red dashed lines, when the two largest GICs in MAN were registered. The mentioned increase is significant for KEV, SOD, OUJ stations located at higher latitudes.  In particular, for the SOD station three peaks around Event I (two before 00:31:20 UT and one after), as well as one peak after Event II, exceed $2000$ mV/km.
For comparison, at a quiet period at 9:00 UT (black dashed line) without large GICs measured in MAN, geoelectric field reveals low  $E$ values (order of a few mV/km) for all stations.

\begin{figure}[!h]
   \centering
\includegraphics[width=0.9\textwidth]{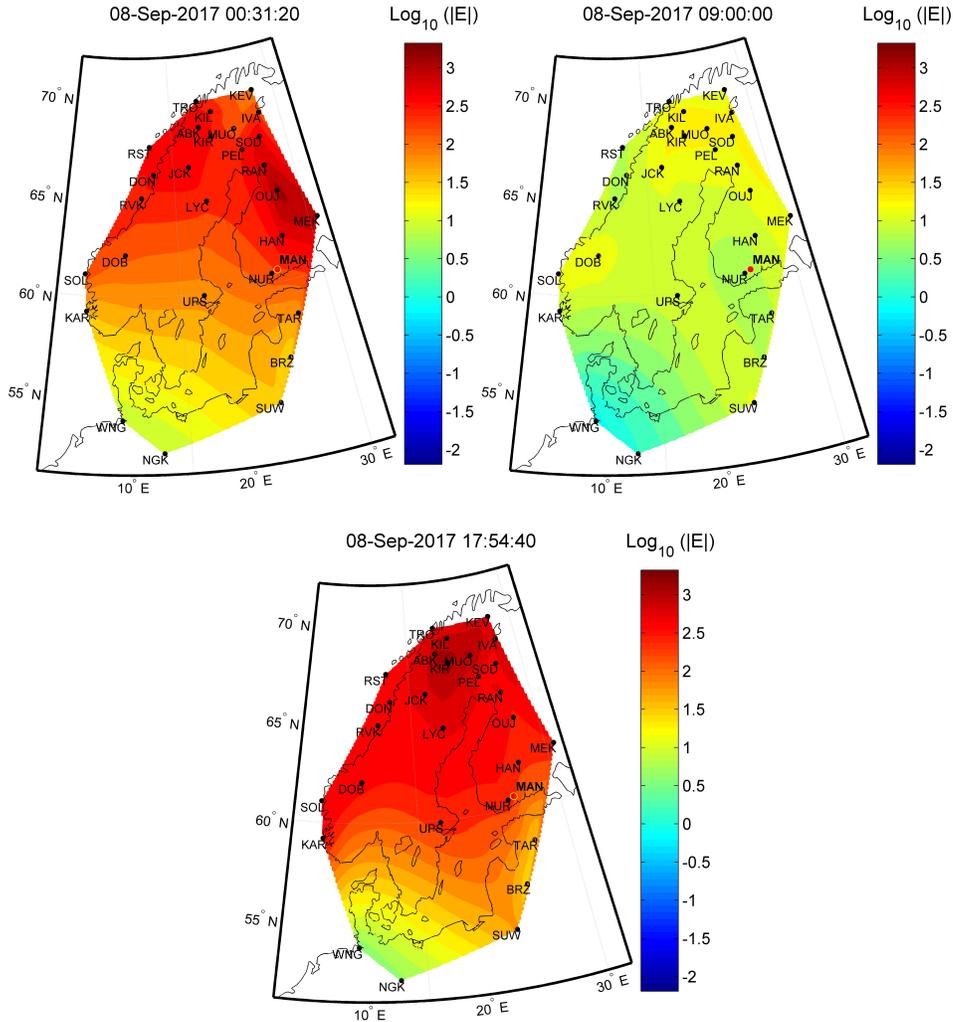}
   \caption{\small The 2-D spatial structure of geoelectric fields $E$ at 00:31:20 UT, 09:00:00 UT, 17:54:40 UT, on 8 September 2017, reconstructed from the IMAGE magnetometers using the 1-D conductivity model. Color coding indicates $\log_{10}(|E|)$ [mV/km].}
   \label{fig:Emap}
   \end{figure}

Figure \ref{fig:Emap} shows a spatial presentation of the magnitude of geoelectric fields $E=\sqrt{E_X^2+E_Y^2}$, for all 28 stations listed in Table \ref{table:1} at three selected moments (which correspond to the dashed lines in Figure \ref{fig:E}): 00:31:20 UT (Event I), 9:00:00 UT and  17:54:40 UT (Event II). Black dots on prepared maps indicate geographic location of the IMAGE magnetometer stations (see Table \ref{table:1}) used for this analysis. For the comparison, the M\"ants\"al\"a station location has been also denoted and highlighted by a red dot. 
Color coding in Figure \ref{fig:Emap} denotes the values of $\log_{10}(|E|)$ [mV/km]. The color scale varies from $|E| = 0.01$ to $3200$ mV/km
 allowing to present a broad range of geoelectric field magnitudes. A magnitude of the field in spaces between stations results from natural neighbor interpolation (see Sect. \ref{sec:4:1}). According to applied procedure, measurements performed by each station in a given time feed interpolation procedure based on which we assign field strength to uniformly distributed locations all over the map. Prepared maps allow to look at the particular event more globally, to consider the spatial distribution of $|E|$, while by use of interpolation procedure one may observe how the region covered by highest signal magnitudes evolves with the time. Moreover, the implementation of the logarithmic scale makes possible observation of the evolution of small signal features what may impose the possibility to catch the development of some effects related to observed phenomena.
Additionally, prepared maps allow to indicate potential values for the geoelectric field over M\"ants\"al\"a, in particular at the time of the GIC peak. Figure \ref{fig:Emap} reveals that in the case of Event I (at 00:31:20 UT) the $|E|$ is on the level 279.7 mV/km ($\log_{10}(|E|)=2.45$ mV/km), while in the case of Event II (at 17:54:40 UT) $|E|$=336.9 mV/km ($\log_{10}(|E|)=2.53$ mV/km).

\begin{figure}[!h]
   \centering
\includegraphics[width=0.9\textwidth]{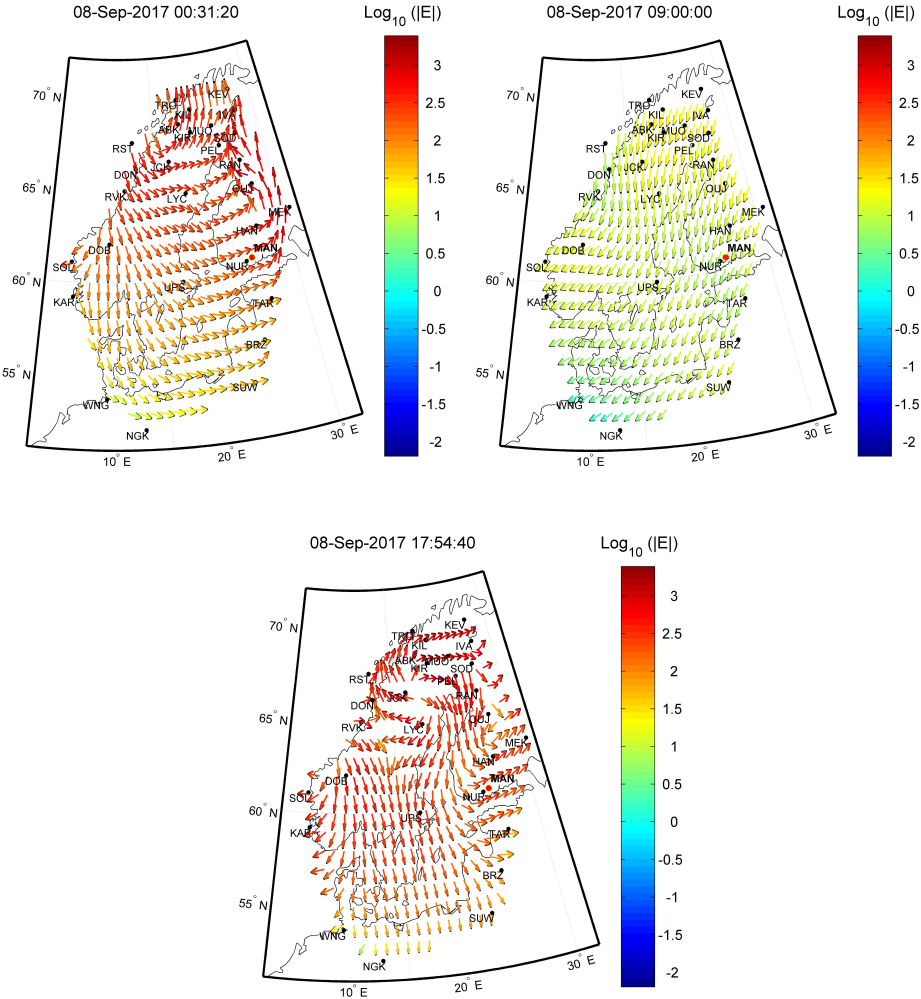}
   \caption{\small The orientation of geoelectric field at 00:31:20 UT, 09:00:00 UT, 17:54:40 UT, on 8 September 2017, reconstructed from the IMAGE magnetometers data. Color coding indicates $\log_{10}(|E|)$ [mV/km].}
   \label{fig:Evector}
   \end{figure}

To identify both local direction and also magnitude of geoelectric field another type of plot has been developed and is shown in Figure \ref{fig:Evector}. Because of the large range of field magnitudes for a single case, we found that a standard approach where the magnitude is coded by the length of arrow is not sufficient. To make image more clear, vectors have been presented by arrows characterized by the same length, appropriate direction and color which represents magnitude ($\log_{10}(|E|)$) of the signal. The size of arrows is unified in a sense of angular distance measured from an origin to the associated tip. This approach is clear, but may result in some graphical effects related to particular projection type. However, it should be underlined that length of an arrow does not contain any relation to any physical signal. 

The orientation of each vector is associated with direction vector calculated for a given origin based on the interpolated $E_X$ and $E_Y$ values for a grid of locations. In this case exactly the same grid is used as in previous plots in this work. For the geoelectric field magnitude presentation predefined range of magnitudes counted as $\sqrt{E_X^2 + E_Y^2}$ has been divided equally, in logarithmic scale, in bins which count is associated with number of colors in particular colormap. In our case, there are 64 different colors in the stack and each color covers a range of magnitudes. The use of a logarithmic scale during the map preparation allows us to present with high accuracy both small values and exceptionally high levels of GEFs. 

\subsection{GIC Analysis}

\begin{figure}[!h]
   \centering
   \includegraphics[width=0.8\textwidth]{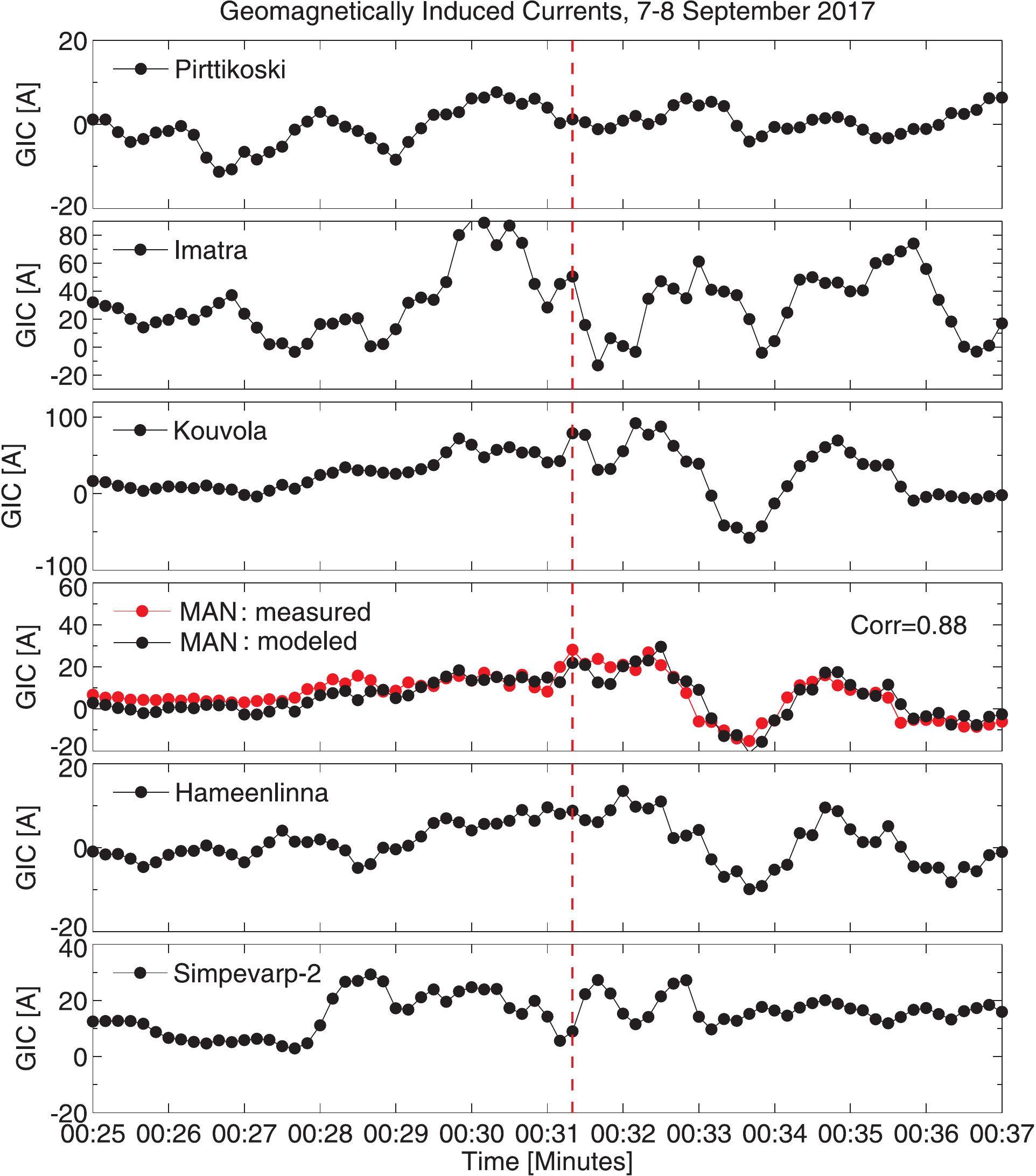}
   \vspace{12pt}
   \caption{\small Time variation of GICs values, determined for period between 00:25 UT and 00:37 UT on September 8, 2017. Results for six selected positions: Pirttikoski, Imatra, Kouvola, M\"ants\"al\"a (MAN), H\"ameenlinna and  Simpevarp-2 listed in Table \ref{table:ab}  have been shown.}
   \label{fig:GICseries_Event1}
   \end{figure}

\begin{figure}[!h]
   \centering
   \includegraphics[width=0.8\textwidth]{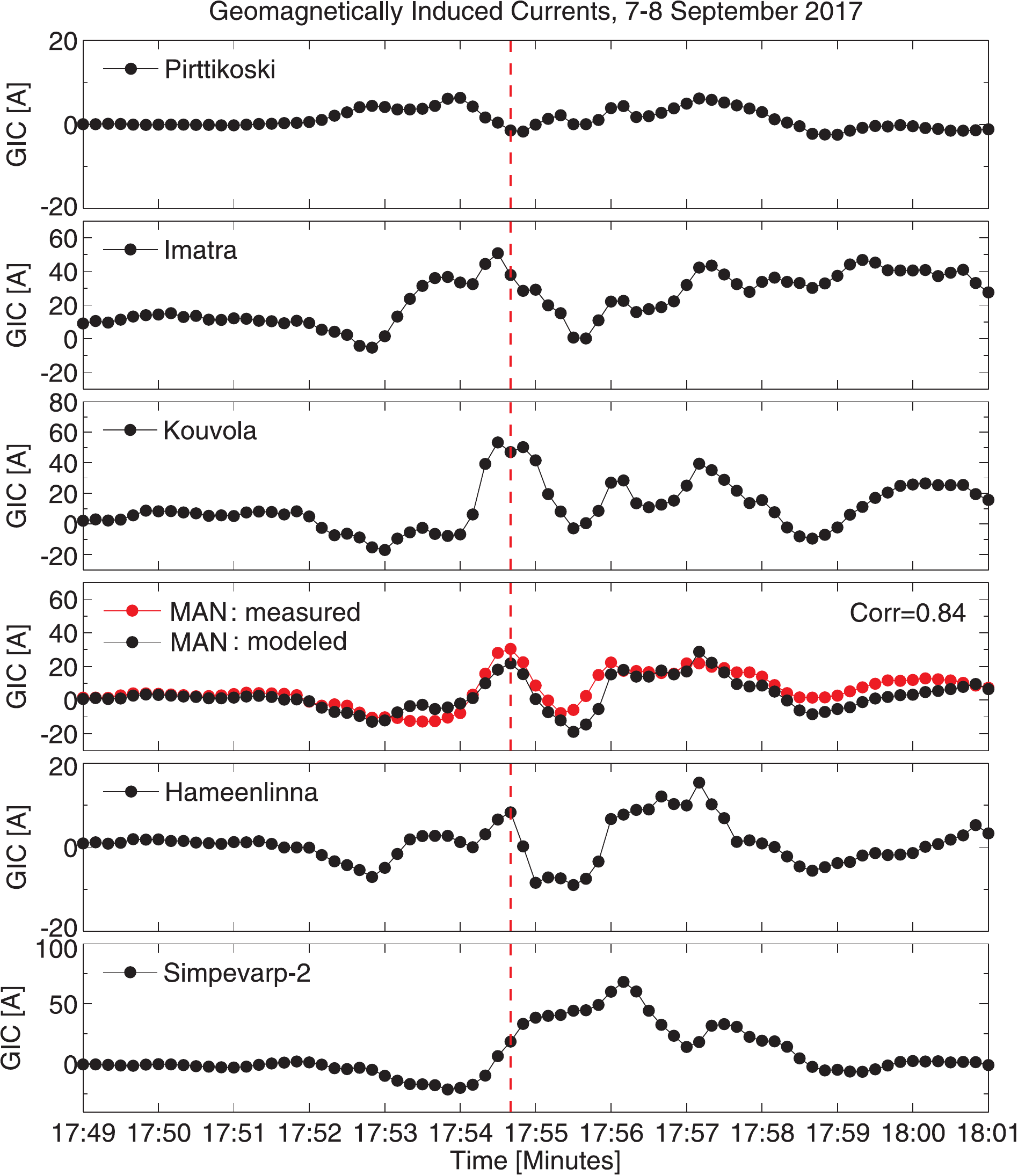}
   \vspace{12pt}
   \caption{\small Time variation of GICs values, determined for period between 17:49 UT  and 18:01 UT on September 8, 2017. Results for six selected  power substations: Pirttikoski, Imatra, Kouvola, M\"ants\"al\"a (MAN), H\"ameenlinna and Simpevarp-2, listed in Table \ref{table:ab} have been shown.}
   \label{fig:GICseries_Event2}
   \end{figure}

 After preparation of 2-D spatial maps of the geoelectric fields $E$, values of GICs were computed (Sect.\ref{sub:m2}). Figures \ref{fig:GICseries_Event1} and \ref{fig:GICseries_Event2} present GICs determined for six selected positions (Pirttikoski, Imatra, Kouvola , M\"ants\"al\"a (MAN), H\"ameenlinna and Simpevarp-2) listed in Table \ref{table:ab}. It is worth emphasizing that the computation of GIC for other substations is outside the scope of this work, as it requires knowledge of grounded conductor network parameters.  Figure \ref{fig:GICseries_Event1} shows the time variation of GICs values for period between 00:25 UT and 00:37 UT on September 8, 2017 (Event I), while Figure \ref{fig:GICseries_Event2} corresponds to period between 17:49 UT and 18:01 UT on September 8, 2017 (Event II). Analysis of Figures \ref{fig:GICseries_Event1}-\ref{fig:GICseries_Event2} reveal the appearance of the strong GICs (around 80 A) at stations Imatra and Kouvola, located in the nearest neighbour to MAN.
 In particular, in the case of Event I the largest GICs appear in station Imatra at 00:30:00 UT, before the moment when GIC event  larger than $20$ A was observed in MAN.  In the case of Event II, the situation is different. The highest values of GICs (around 50A) at stations Imatra, Kouvola are observed exactly at the same moment like for MAN, namely at 17:54:40 UT. For stations located at lower latitudes, H\"ameenlinna and Simpevarp-2 the higher GICs appear after mentioned moment.
 
 Our modelling is able to reproduce the GICs measurements in MAN, with the correlation coefficient of $\sim 0.88$ (Event I) and $\sim 0.84$ (Event II) for measured (red line) and modelled (black line).
 GIC values during the geomagnetic storms are usually in the order of tens
 of amperes \cite<e.g.>[]{2021Svanda,2021Torta} and our results confirms these observation (we do not observe values higher than 100 A).  Moreover, GICs computations correspond to values mentioned for mid-latitudes by \cite<e.g.>[]{Albert2022,Bailey2022}.
For comparison, for Simpevarp-2 power substation on 06 April 2000, \citeA{W2008} reported the highest GIC detected on a power transmission line: 300 A. The highest calculated value of GIC at Simpevarp-1 was found at the level of 330 A during the Halloween Storm, on 30 October 2003, around 20:00 UT \cite{W2009}.
 Our estimated peak amplitudes for the September 2017 storm are comparable to a similar analysis performed for neighboring Austria. More precisely, \citeA{Bailey17} estimated the GIC amplitudes of about 13 A.
 
 It is worth underlining another observation. The difference in spatial variability between induced electric fields and computed at selected points GICs demonstrates the significant influence of the network parameters $a$ and $b$ on the level of the  GICs. 

\subsection{Geoelectric Field During Event I}
\label{sub:Event I}

\begin{figure}[!ht]
\centering
\includegraphics[width=0.9\textwidth]{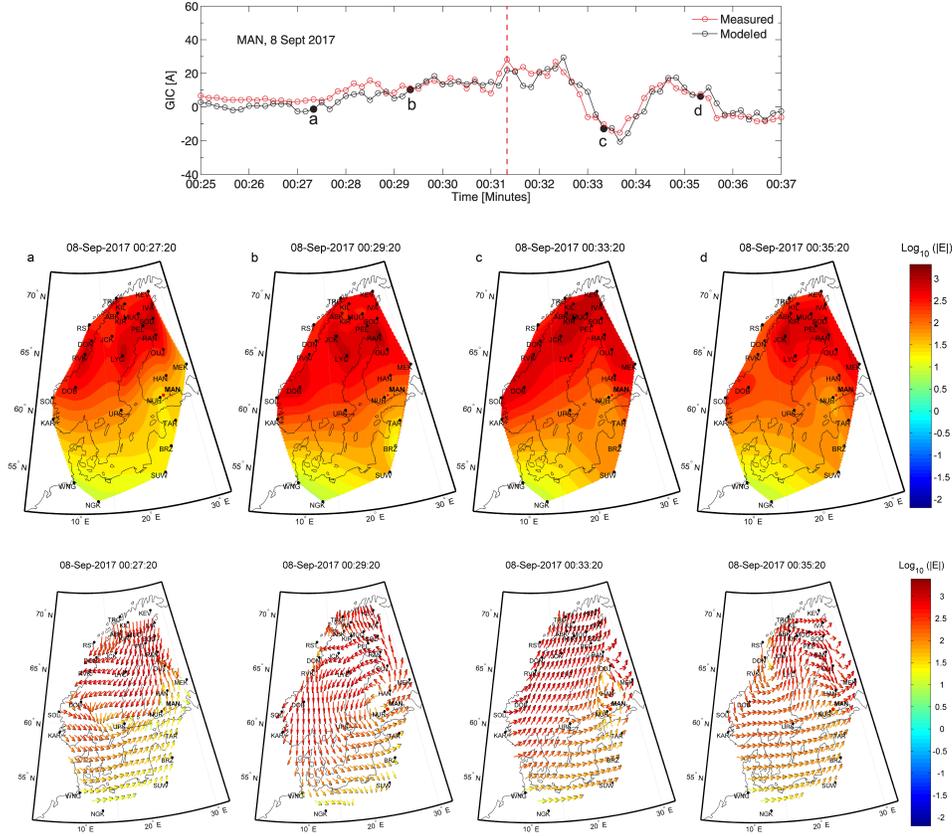}
\caption{Measurement of GICs in MAN (red line) compared with modelled in the frame of this paper (black line) for period when Event I occurred at 00:31:20 UT (red dashed line). Additionally, four moments: a (Event I-4min), b (Event I-2min), c (Event I+2min), and d (Event I+4min), have been indicated and maps of $E$ magnitude in mV/km  (top) and $E$ direction (bottom) have been presented.}
\label{fig:event1}
\end{figure}

Animation of maps present the time and spatial spread of computed geoelectric field $E$ of few minutes around Event I of September 8, from 00:25 to 00:37 and the corresponding GIC in M\"ants\"al\"a (please, see Movie S1 in Supporting Information).\\  
The first stage of the animation from 00:25 to 00:27 shows the quasi stable conditions of GIC measurements at M\"ants\"al\"a demonstrated as a flat line of GIC about 5 A,  with enhancements developing in North. Next around the time of 00:27:40 to 00:28:40 one can see the development of dark structures of large $E$ to the left hand side, covering the area around the DOB, RVK, DON, JCK and LYC stations. It was also seen at M\"ants\"al\"a location at 00:28:30 as a local maximum of GIC. Afterward one can observe the moving dark structures to the right, covering the area around the SOD, PEL, RAN, OUJ, HAN and MEK stations and reaching the peak value of GIC around M\"ants\"al\"a location at 00:31:20 (please see left top panel of Figures \ref{fig:Emap} and \ref{fig:Evector}).  The highest value of modeled geoelectric field there was at 00:32:30 reaching 360.5 mV/km. Next, values of GIC around M\"ants\"al\"a start decreasing and afterwards changing sign, seen as local minimum between 00:33 and 00:34 which is in agreement with computed geoelectric field. Next the situation develops dynamically starting at 00:34:00 lasting  for about 1 minute, with the strong rise at 00:34:40 observed at M\"ants\"al\"a location.  After that,  starting at 00:35:40 one can see the quasi stable situation observed at M\"ants\"al\"a location connected with local decreasing of computed $E$ around this region.\\
It is well known, \cite{16Alberti, 94Araki, 16Araki, 16Piersanti, 17Piersanti, 09Villante}, that at low latitudes, the contribution of the magnetospheric origin is dominant. Whereas, for higher latitudes, the contribution of the ionospheric origin is dominant \cite{16Alberti, 94Araki, 16Piersanti}  as also shown by IMAGE data in \cite{Dimmock2019}. The selected maps of the magnitude and direction of the geoelectric field $E$  for the whole area of IMAGE data  for the same periods as refered , in \cite{Dimmock2019}, particularly for the time of Event I (left top panel of Figures \ref{fig:Emap} and \ref{fig:Evector}), 2 and 4 minutes before and after the Event I are presented in Figure \ref{fig:event1}. The upper panel of Figure \ref{fig:event1} presents a comparison of the GIC measurements in M\"ants\"al\"a and modeling results read from maps. One can see that the time profiles of modelled and measured GICs in M\"ants\"al\"a are in quite acceptable agreement. At 00:31:20, when the measured values reached its peak the modelled values are somewhat underestimated, with maximum at 21.84 A.\\
Four minutes before the GIC peak the large geoelectric field $E$ is observed in the range of latitudes $\sim60-70^{o}$, which is consistent with and ionospheric current and the extension of the auroral oval towards the equator during the geomagnetic storm. Two minutes later the band splits up and more structures can be found. Therefore, the same large-scale west-east current is observed at this time, but there seems to be an effect of smaller spatial-temporal structures that overlap or are embedded in a larger system of currents. It is found that the electrojet has structure in the form of rapid and strong changes in the level and orientation of the geoelectric field. Even though, the dynamics of the localized structures around Event I is rather complex, one can see quite good correspondence between maps of ground modelled geoelectric field $E$ magnitudes for IMAGE data and ionospheric currents presented in \cite{Dimmock2019}. Dynamics shows evidence of many sources in different scales, both in time and space. This is very important from the point of view of the regional behavior of geoelectric field $E$ realized in this paper and hence potential GIC risk.

\subsection{Geoelectric Field During Event II}
\label{sub:Event II}
\begin{figure}[!ht]
\centering
\includegraphics[width=0.9\textwidth]{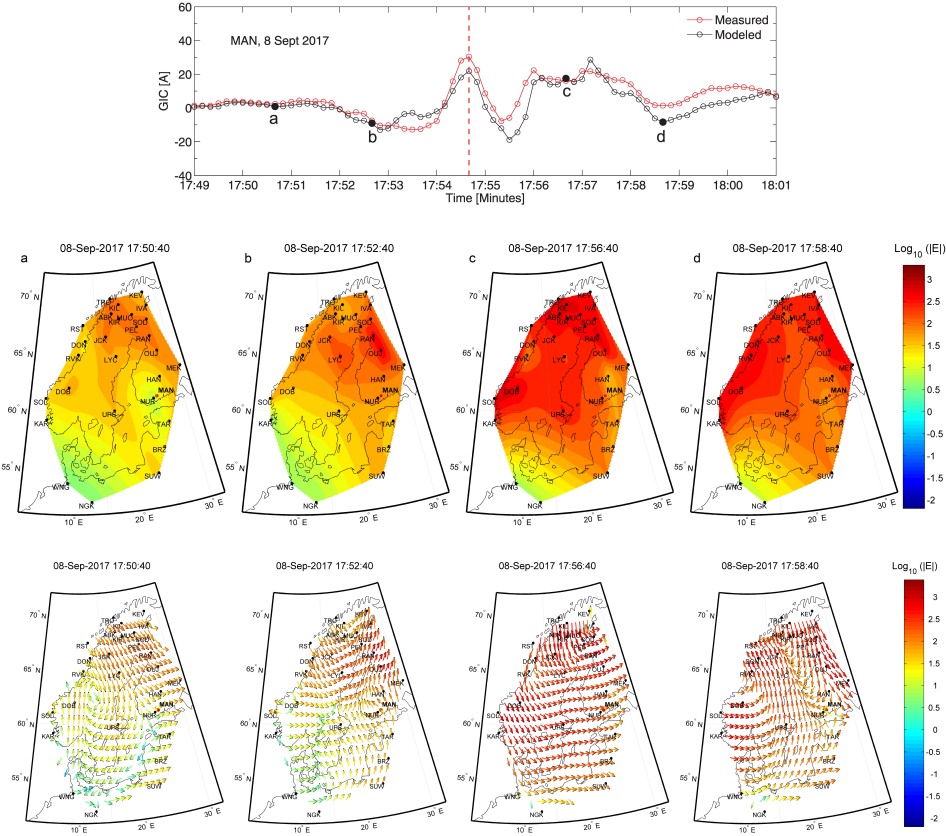}
\caption{Measurement of GICs in MAN (red line) compared with modelled in the frame of this paper (black line) for period when Event II occurred at 17:54:40 UT (red dashed line). Additionally, four moments: a (Event II-4min), b (Event II-2min), c (Event II+2min), and d (Event II+4min) have been indicated and maps of $E$ magnitude in mV/km (top) and $E$ direction (bottom) have been presented.}
\label{fig:event2}
\end{figure}

Maps showing the spatial-temporal variability of the horizontal geoelectric field values (Figure \ref{fig:Emap}, lower panel, and Movie S2 in the Supporting Information, left panel) show that stations located at higher latitudes were characterized by an increased value of $E$ already a few minutes before Event II. In the vicinity of M\"ants\"al\"a, the situation developed dynamically until exactly 17:54:40, i.e. the occurrence of the highest measured GIC value. This high $E$ value persisted in the vicinity of M\"ants\"al\"a for about 20 s. The highest value of modeled geoelectric field was at 17:54:50 reaching level of 396.0 mV/km. 
After that it gradually began to weaken. At 17:57:10 it again reached the highest levels for this location (310.9 mV/km), which was reflected in the GIC measurements, namely at the same time when there was an increase above 20 A. However, it should be mentioned that also at 17:56:00  a rise above 20 A was recorded in M\"ants\"al\"a, not linked to the increase in the value of $E$ shown on the map.\\
The upper panel of Figure \ref{fig:event2} presents the chronological evolution of computed GIC in comparison with the GIC measurements in M\"ants\"al\"a. One may see a good agreement between these two. At 17:54:40, when the measured values reached its peak, the modeled values are a bit underestimated, with maximum at 21.84 A.
In the modeling results there is also a strong peak that is only slightly less powerful in the measurements, at 17:57:10, being 28.70 A.
Since the basis for determining both $E$ and GIC is $dB/dt$, therefore a comparison with the measurements of the closest magnetometer in Nurmij\"arvi (NUR), reveals that the moment of the highest GIC value observation coincides with the moment of the strongest fluctuations in the local geomagnetic field. Comparing with the second closest station, i.e. the more distant Hankasalmi (HAN), located farther north, expose that the highest GIC registration in M\"ants\"al\"a was 20 s after local maximum appeared in $dB/dt$ in Hankasalmi observations.\\ 
Analysis of geoelectric field vector behavior during the Event II (the lowest left panel of Fig. \ref{fig:event2}) display that 4 minutes before the GIC peak appeared, $E$ around M\"ants\"al\"a was eastwardly directed. The highest values appeared around north-eastern part of the considered region (the middle panel of Fig. \ref{fig:event2}). In the next two minutes these enhancements expanded more towards the east and around M\"ants\"al\"a vector $E$ was directed to north-west. During the moment of the highest GIC measurements (Figure \ref{fig:Evector}, lower panel) practically the whole studied region was covered by the amplified $E$ field. There were no clear organisation in the vector direction. Around M\"ants\"al\"a geoelectriec field vector was again directed to east.
This situations developed dynamically and two minutes later the $E$ strengthening extended even more to the whole region, with some exceptions at the south and north ends. Vector field was presenting an ordered situation with its direction around M\"ants\"al\"a being still east. Some reductions of geoelectric field values was visible within next two minutes with strong disorganization in its direction. \\
It is worth noting, that for the northern sites the Event I was stronger in the geoelectrical field changes and in GIC intensity (compare Figures \ref{fig:Emap} and \ref{fig:GICseries_Event1}). Although for more mid-latitudes sites the Event II was characterised by more forceful variations and growths, both in $E$ and GIC  values (compare Figure \ref{fig:Emap} and last three panels of Figure \ref{fig:GICseries_Event2}). \\
Animations included in the Supporting Information showing the spatial-temporal development of the calculated local geoelectric field and the computed GIC with GIC measurements in M\"ants\"al\"a, simultaneously, allow us to observe that the values exceeding the maxima for the Event I and II appeared several times, in different locations, not only in the northernmost ones. They also show that rather stable structures are moving up and down, being a possible cause of the registered GIC values around 30 A. Moreover, a movement of an area characterised by the increased $E$ values (shown as dark red area) across one station might indicate auroral expansion and associated electrojet. Thus, one may state that the electrojet has variations in a form of swift change in the level and orientation of the geoelectric field. But some of these suddenly appearing higher values (visible as darkest areas) might also mean strong precipitation down local magnetic field lines.\\
Comparing with ionospheric equivalent currents presented in \cite<Fig. 10 in>[]{Dimmock2019} for the same time intervals we should not expect one-to-one correspondence since here we present modelled geoelectric field at lower latitudes. Our results show massive structural evolution evolving in time, similar to \citeA{Dimmock2019}. We both show strengthening of parameter values at similar locations.

\subsection{Discussion}

Knowledge about the geoelectric field $E$  remains crucial due to being the main driver for creating GICs in conductor networks on the surface of the Earth.
Geoelectric field computation performed in this work is in an agreement with  measurements maintained in Great Britain \cite{Beget21}. Namely, the geoelectric field variation in Great Britain during this substorm was over 1 V/km in Lerwick, 0.5 V/km in Eskdalemuir and 80 mV/km in Hartland  \cite<Figure 3 in>[]{Beget21}. Moreover, our calculations of $E$ for ABK and UPS stations are comparable with the \citeA{22Krug} computation of the geoelectric field, for real-time, using the SECS method applied to the geomagnetic field from the IMAGE network.  
Moreover, geoelectric field mapping from Sect.\ref{sec:4}, in particular the regions with strong values of $E$, denoted by red dark color, state independent confirmation of previous observations that the North Europe is the most likely area of large geoelectric fields \cite<e.g.>[]{Vilet14}. On the other hand, we see that significant geoelectric field disturbances may also occur at much lower latitudes, indicating the possibility of GIC problems there too \cite{Beget21, Gilet21, 2020Piersanti, 2021Svanda, 2021Torta}. 
When the situation on the maps presented in Sect.\ref{sec:4} and in the form of the Supplementary Information is compared with GIC measurements in M\"ants\"al\"a, a good qualitative agreement is apparent. 
Referring to the detailed discussion about particular events (see Sect. \ref{sub:Event I} and \ref{sub:Event II}) it is worth pointing out a few observations. 
First of all, our study suggests that the northern sites during the Event I were characterized by stronger variations in the geoelectric field (Figure \ref{fig:E}-\ref{fig:Emap} and \ref{fig:GICseries_Event1}). However, for the more mid-latitudes IMAGE observatories the Event II presents more powerful changes variations and growths, both in $E$ and GIC values (Figure \ref{fig:E}-\ref{fig:Emap} and \ref{fig:GICseries_Event2}). In particular, our results reveal a higher level of geoelectric fields over M\"ants\"al\"a at the time of the peak GIC for Event II ($|E|$=336.9 mV/km) than for Event I ($|E|$=279.7 mV/km).
Moreover, the Supplementary Information displaying the spatial-temporal development of the computed local geoelectric field and GIC values with GIC measurements in M\"ants\"al\"a, all together, let us detect that the values beyond the maxima for the Event I and II occurred numerous times, at various sites, not only the northernmost one. The animations also present rather stable structures moving up and down for Event II and from left to the right for Event I, being a potential reason of the GIC 
appearance of about 30 A. Furthermore, a movement of a dark red area across one station might indicate auroral expansion and with it the electrojet. It is found that the electrojet characterizes structure in a form of the geoelectric field rapid change, both in the level and orientation. But some of these abruptly occurring increased $E$ values, shown as the darkest area, might also mean strong precipitation down local magnetic field lines.
The question arises what will be the picture of the evolution and change of the $E$ field when the lower resolution is applied. \citeA{Pulet06} stated that the 1 min sampling rate of the ground magnetic field is able to capture essentially the same features of the surface geoelectric field variations as that of the 1 s sampling rate. 
Our analyses of GIC measurements in M\"ants\"al\"a (the first panels of Fig. \ref{fig:event1} and \ref{fig:event2}) and the comparison of $E$ maps and GIC values for two resolutions 10 s and 1 min (not shown here) suggested that changes of geoelectric field are very rapid, and to analyze them correctly temporal resolution shorter than 1 min is required. Applying 10 s time resolution data we were able to observe on the maps how high $E$ fluctuations appear, evolve both spatially and temporarily, as well as fading away. This consequent behavior of phenomena made us sure that we are analyzing data on appropriate scales. \\
Our finding about  the resolution is in agreement with recent papers \cite<e.g.>[]{Rodet17, Tri21}, which recommended the consideration of high temporal resolution of the used geomagnetic data (seconds, to tens of seconds). In particular, \citeA{Ganet17} concluded that geoelectric field calculation based on 1 min data introduced a loss of 50\% of its peak value (in comparison with 10 s data) and underlined that resolving the frequency content may be more important than accurate 3-D modeling of the Earth response.

\section{Conclusions}
\label{sec:5}
In this article, we have performed systematic computation and consideration of spatio-temporal variability of the geoelectric field (both magnitude and direction) during  7--8 September 2017 geomagnetic storm - one of the largest storms of the 24th solar cycle. More precisely, we have focused on two largest GICs events (Event I and II) registered in Finnish natural gas pipeline near M\"ants\"al\"a. 
Since $B$ field measurements were generally not available at the locations of interest for calculation of the $E$ field we have developed the GeoElectric Dynamic Mapping (GEDMap). GEDMap results revealed the directional evolution of structures, a rapid change of the orientation of geoelectric field, as well as differentiated situation for Event I and Event II.\\ 
More precisely, Event I was characterized by enhanced geomagnetic activity, but the largest GIC observed in M\"ants\"al\"a occurred as Event II, when the geomagnetic enhancements were not of their highest amplitude, as was mentioned in \cite{Dimmock2019}.
Although, it is difficult to associate the Event II with any large substorm or any solar wind feature \cite{Tsurutani2021}, our mapping seems to be a very useful tool for observation of temporal and spatial movements. Presented maps showed rather stable dark structures moving up and down for Event II and from left to the right for Event I. Additionally, our results suggested a higher level of geoelectric fields over M\"ants\"al\"a at the time of the GIC peak for Event II than for Event I. Moreover, in some intervals computed geoelectric fields  around M\"ants\"al\"a region were changing sign, being in agreement with GICs observations.\\
Summarizing, geoelectric field mapping prepared within this work give a global perspective and state a useful tool for simultaneous observations of spatial and temporal variation of considered quantities. The possibility of the observation of their changes for many stations in the same time, to catch how some spatial structures develop, move and spread can be interpreted and coupled in the future with various phenomena \cite<e.g.>[]{Oliet18}. \\
Our work is ongoing, and future studies will focus on analysis of other important events, in particular on Halloween storms in 2003, where very large GICs appeared \cite{Hajra2020}. As \citeA{Tsurutani2021} suggested the interplanetary and magnetospheric causes of GICs at auroral latitudes, mid-latitudes, and equatorial latitudes might be different. It would be interesting to apply GEDMap in this context.\\
Moreover, the GIC problems at middle- and low-latitude transmission lines have increasingly gained attention \cite<e.g.>[]{2021Svanda, Bailey2022, Albert2022, 19Tozzi, 13Zois, 12Torta, 15Barbosa}. In the light of this task proposed procedure of the geoelectric field  mapping, not based on SECS methods (appropriate for high latitudes) can be tested and applied.\\
Next, our results showed that geoelectric field mapping with 10s scales give a lot of details worth of systematic consideration/interpretation, while identified changes move. It will be interesting to check the possibility of application even higher 1s resolution, as recommended by \cite{Tri21},  and the verification how it influences the results.\\
The forthcoming study will also take into account the verification of other Earth modeling assumptions.
The modern tendency is to use more realistic, 3-D models of surface impedance \cite<e.g.>[and references therein]{Ganet17, Kleet17, KelLuc20}, in particular some works suggest that two-dimensional and three-dimensional Earth conductivity structure introduces some features not seen with 1-D models \cite<e.g.>[]{BotPir17}. Nevertheless, it is worthwhile to mention that a 1-D model is still treated as accurate at a single location \cite{Beget21}. 
 We are convinced that further systematic and detailed study of proposed geoelectric field  mapping will be important for better understanding causes of GICs and shed more light for these complex geophysical phenomena.
 
 \section*{Open Research}
Data of geomagnetic field  components and geomagnetic indices are from IMAGE - International Monitor for Auroral Geomagnetic Effect, \url{http://space.fmi.fi/image/}. GIC recordings in the Finnish natural gas pipeline near M\"ants\"al\"a are from Finnish Meteorological Institute, \url{https://space.fmi.fi/gic/}.

\acknowledgments
Portions of this research were performed at the Jet Propulsion Laboratory, California Institute of Technology under contract with NASA.


%
\bibliography{spaceweatherAW} 
%




\end{document}